The paper will be presented at the *2027 21th European Conference on Antennas and Propagation (EuCAP)*, Düsseldorf, Germany, April 18–23, 2027

# Sensitivity of RMS Delay Spread Estimation to Power Delay Profile Thresholding in Time-Varying Channels at 33 GHz

Jarosław Wojtuń[1], Bartosz Czaja[1], Tomasz Grasza[1], Jan M. Kelner[1], Cezary Ziółkowski[1], Radek Zavorka[2], Josef Vychodil[2], Tomas Mikulasek[2], Jiri Blumenstein[2], Ales Prokes[2], and Aniruddha Chandra[3]
[1] Institute of Communications Systems, Faculty of Electronics, Military University of Technology, Warsaw, Poland, *jaroslaw.wojtun@wat.edu.pl*
[2] Department of Radio Electronics, Brno University of Technology, Brno, Czech Republic
[3] ECE Department, National Institute of Technology, Durgapur, India[1]

***Abstract*—Root mean square (RMS) delay spread is a key parameter characterizing temporal dispersion in radio propagation channels. Its estimation from measured power delay profiles (PDPs), however, depends strongly on the adopted power threshold. This paper investigates the sensitivity of RMS delay spread estimates to relative PDP thresholding in a time-varying millimeter-wave channel at 33 GHz. The analysis uses wideband channel sounding measurements conducted in a dynamic road environment with moving vehicles. The temporal evolution and empirical cumulative distributions of RMS delay spread are evaluated for thresholds ranging from −20 to −50 dB. Reducing the threshold increases the median RMS delay spread from approximately 140 to 168 ns and the 90th percentile from 200 to 210 ns. Both statistics converge at lower thresholds, with −40 dB providing a practical convergence level. The results emphasize the importance of threshold selection for reliable channel characterization and consistent comparison of measurements.**

## I. Introduction

Temporal dispersion is one of the fundamental characteristics of radio propagation channels, describing the spreading of transmitted signals in the delay domain due to multipath propagation. Its characterization is particularly important in time-varying channels, where the motion of transmitters, receivers, or surrounding objects causes continuous changes in the multipath structure. The power delay profile (PDP) is a fundamental representation of the received signal power distribution over propagation delay. Based on the PDP, the mean delay and root mean square (RMS) delay spread can be determined as the first moment and the square root of the second central moment of the delay distribution, respectively [1], [2]. Among these parameters, RMS delay spread is particularly important for assessing frequency selectivity, intersymbol interference, and the performance of wideband communication systems. It is also widely used in radio channel modeling and in distinguishing propagation environments according to their temporal dispersion characteristics [1].

For a measured discrete PDP, the RMS delay spread, $\sigma_\tau$, is defined as [1], [3]

$$\sigma_\tau = \sqrt{\frac{1}{\overline{P}}\sum_{k=1}^{K} P_k\left(\tau_k - \overline{\tau}\right)^2} \qquad (1)$$

where $\overline{P} = \sum_{k=1}^{K} P_k$ denotes the total power of the retained PDP samples and $\overline{\tau} = \frac{1}{\overline{P}}\sum_{k=1}^{K} P_k\tau_k$ represents the mean delay, while $P_k$ and $\tau_k$ denote the power and propagation delay of the $k$th PDP sample, respectively. The summations include only samples satisfying the adopted thresholding criterion. Consequently, the estimated RMS delay spread depends not only on the physical multipath characteristics but also on the selection of the PDP power threshold.

In the literature, different threshold definitions and terminology are employed. The *noise threshold*, also referred to as the *noise floor threshold* or *noise exclusion threshold*, is established above the estimated noise floor to suppress noise-related components [2], [4], [5]. In contrast, the *multipath threshold*, also termed the *quality threshold* in [4], is defined relative to the strongest PDP component. In this paper, the latter is referred to as the *relative PDP power threshold* or *cutoff threshold*, denoted by $Th$. Its value is expressed in decibels relative to the peak power of each individual PDP. To avoid including noise-dominated samples, the resulting cutoff level should remain above the estimated noise threshold.

The influence of threshold selection on RMS delay spread estimation has been addressed in several experimental studies. Pomianek *et al.* [6] demonstrated that the estimated RMS delay spread depends strongly on the considered PDP dynamic range and approaches a stable value when sufficiently weak multipath components are included. Rubio *et al.* [7] investigated time-dispersion characteristics of UWB channels for different threshold levels, while Martinez-Ingles *et al.* [8] reported noticeable differences in RMS delay spread

This research was funded in part by the National Science Center (NCN), Poland, grant no. 2021/43/I/ST7/03294 (MubaMilWave).

estimates obtained using 20 and 30 dB thresholds in centimeter- and millimeter-wave (mmWave) measurements. Coulibaly *et al.* [5] compared fixed-level and dynamic noise removal methods, demonstrating their impact on RMS delay spread and other channel parameters. More recently, Schulpen *et al.* [4] systematically investigated ambiguities in RMS delay spread estimation from mmWave measurements, distinguishing between noise and multipath thresholds and recommending a combined thresholding approach to improve the consistency of results across measurement campaigns.

Despite these contributions, the sensitivity of RMS delay spread statistics to relative PDP power thresholding in strongly dispersive, time-varying mmWave channels remains an important issue for experimental characterization. This paper investigates this problem using wideband channel sounding measurements conducted at 33 GHz in a dynamic road environment. The considered quasi-monostatic scenario includes stationary scatterers and moving vehicles, producing pronounced variations in the multipath delay structure. The analysis focuses on the influence of PDP cutoff thresholds ranging from −20 to −50 dB on the temporal evolution and empirical cumulative distributions of RMS delay spread. Particular attention is given to the median and 90th percentile (P90) and to identifying the threshold range beyond which these statistics exhibit negligible changes. The main contribution is an empirical assessment of the sensitivity and statistical convergence of RMS delay spread estimates to relative PDP thresholding in a vehicle-induced, time-varying mmWave propagation environment at 33 GHz, providing additional guidance for threshold selection and consistent comparison of temporal dispersion measurements.

The remainder of this paper is organized as follows. Section II describes the measurement setup and the investigated propagation scenario. Section III presents the measured PDPs and analyzes the influence of threshold selection on RMS delay spread and its statistical characteristics. Finally, Section IV summarizes the main findings and conclusions.

## II. Measurement Campaign

### A. Measurement Setup

The empirical investigation was conducted using a wideband millimeter-wave channel sounding system illustrated in Fig. 1. The testbed, previously employed in related propagation studies [3], [9], comprises separate transmitting (Tx) and receiving (Rx) units based on a Keysight M9384B VXG vector signal generator and a Keysight N9032B PXA signal analyzer, respectively. Both instruments operate at frequencies up to 44 GHz and support instantaneous measurement bandwidths (BWs) of up to 1 GHz.

The Tx chain incorporates a Ka-band power amplifier (PA) (Eravant SBP-2034434526-2F2F-S1), providing 45 dB gain over the 20–44 GHz frequency range. The signal generator is connected to the PA through a 5 m low-loss coaxial cable, while a 0.5 m feeder connects the amplifier output to the Tx antenna. The Rx chain consists of an Rx antenna, a Ka-band low-noise amplifier (LNA) (Eravant SBL-2434434025-2F2F-E3) with a nominal gain of 40 dB over the 24.25–43.5 GHz frequency range, and the signal analyzer connected to a computer for data acquisition and storage. Similarly to the Tx configuration, 0.5 m and 5 m low-loss coaxial cables connect the Rx antenna to the LNA and the LNA to the analyzer, respectively. Time and frequency synchronization between the Tx and Rx instruments is provided by a Stanford Research Systems FS740 GPS-disciplined frequency and time reference.

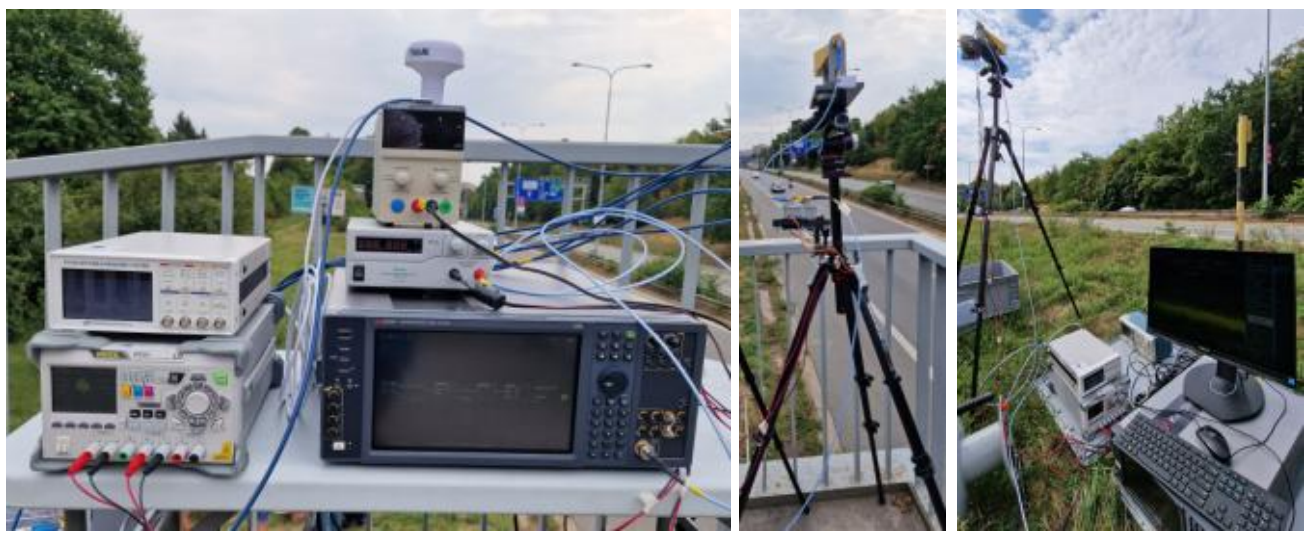

(a) (b) (c)

Fig. 1. Measurement testbed: (a) TX side, (b) TX antenna, (c) RX side.

Identical Eravant SAR-2507-28-S2 rectangular horn antennas were employed at both ends of the measurement link. These antennas cover the 26.5–40 GHz frequency band, offer a nominal gain of 25 dBi, and exhibit half-power beamwidths (HPBWs) of approximately 7° and 9° in the elevation and azimuth planes, respectively.

Channel sounding was performed using a frequency-modulated continuous-wave (FMCW) chirp signal with a bandwidth of 1 GHz and a duration of 8 µs. The sampling rate of both the Tx and Rx chains was set to 1.2 GSa/s. On the Rx side, data acquisition was controlled by an external function generator Tektronix AFG3021, which triggered the signal analyzer to capture 16 µs recording windows at intervals of 25 ms, corresponding to a recording rate of 40 Hz. Each acquisition window contained at least one complete FMCW chirp suitable for subsequent channel analysis. The total measurement duration was 30 s, resulting in 1,200 recording windows and approximately 1,200 complete chirps available for processing. All measurements presented in this study were conducted at a carrier frequency of 33 GHz.

### B. Measurement Scenario

The measurement campaign was carried out on August 26, 2026, along Hradecká Street (road 640/E461), near the campus of the Brno University of Technology (BUT), Brno, Czech Republic. An aerial view of the investigated road section and the deployment of the measurement equipment is presented in Fig. 2.

The Tx unit was installed on a pedestrian footbridge crossing Hradecká Street and connecting Technická and Purkyňova Streets. The Rx unit was positioned near the roadside, slightly ahead of the Tx location along the monitored road section. The transmitting and receiving antennas were mounted approximately 5 m and 1.5 m above road level, respectively.

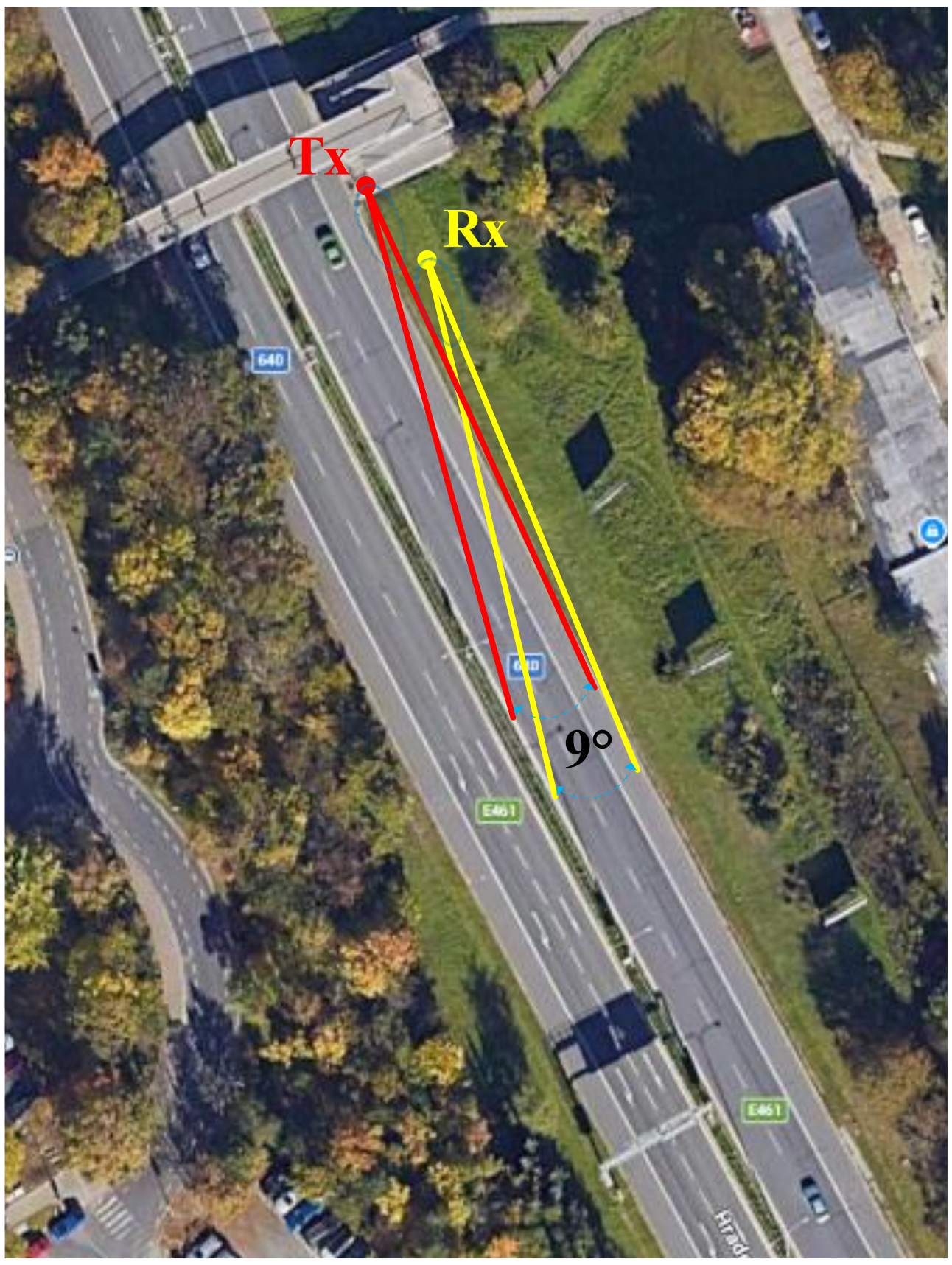

Fig. 2. Aerial view of measurement scenario along Hradecká Street near BUT campus, showing Tx and Rx locations and monitored traffic direction (background map: Google Maps).

Both directional antennas were oriented approximately along the same direction, pointing toward the nearest traffic lane and facing oncoming vehicles. Due to the relatively small separation between the Tx and Rx antennas compared with the distances to the observed vehicles, the adopted geometry can be regarded as a quasi-monostatic configuration, resembling that commonly employed in monostatic radar systems. The direct Tx–Rx propagation path was outside the main antenna beams, and the investigated vehicle-related propagation components were therefore considered under non-line-of-sight (NLOS) conditions.

The surrounding environment was predominantly open, with relatively few stationary scattering objects. The main static contributors to multipath propagation included street lighting poles, traffic signs, overhead road information boards, and metal safety barriers separating opposing traffic directions. In contrast, vehicles passing through the illuminated road section introduced time-varying scattering and reflection components, resulting in a dynamically evolving multipath channel. Depending on the instantaneous traffic conditions, different numbers and types of vehicles contributed to the received signal, producing variations in the PDP and the associated temporal dispersion characteristics.

## III. Results

Fig. 3 illustrates the temporal evolution of the PDP obtained from a representative 30 s measurement record at 33 GHz. Individual channel impulse responses (CIRs) were estimated by correlating the received FMCW signals with the reference waveform. To suppress stationary multipath components, second-order temporal differencing with finite impulse response (FIR) coefficients [1, −2, 1] was applied to consecutive complex CIRs. The corresponding PDPs were then calculated as the squared magnitudes of the filtered CIRs.

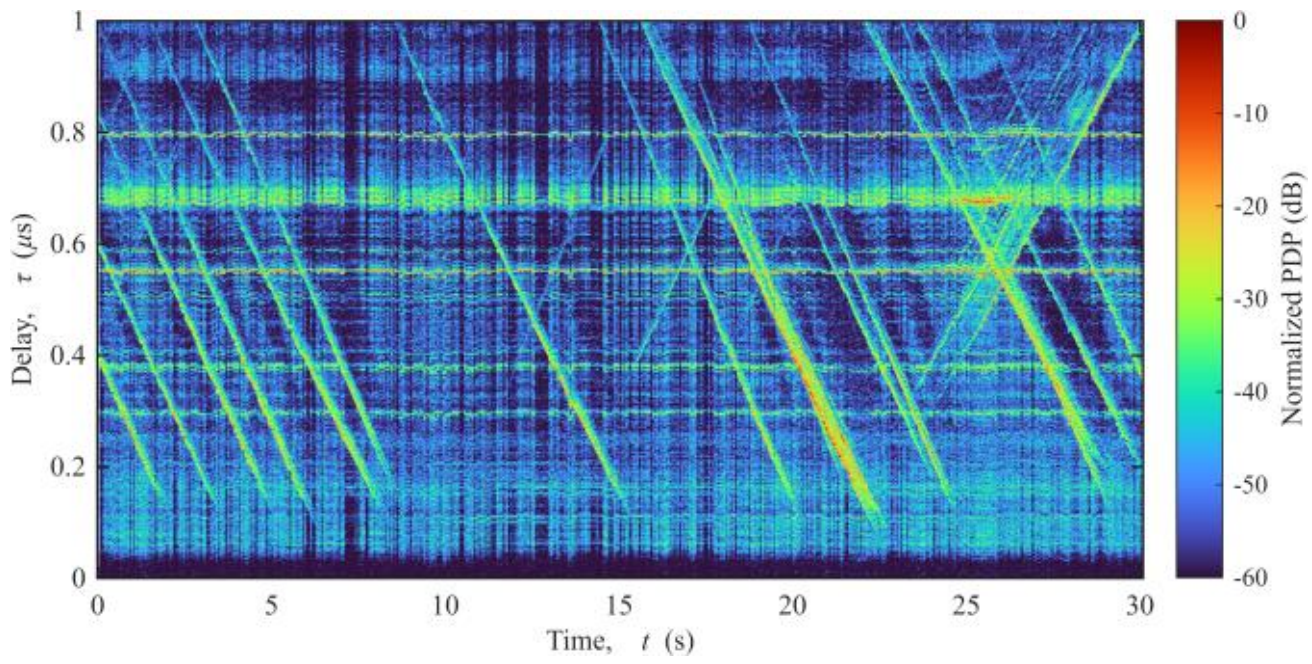

Fig. 3. Temporal evolution of measured PDP at 33 GHz after second-order FIR filtering.

The resulting time-delay representation reveals both stationary and time-varying multipath components. The remaining nearly horizontal traces may be attributed to residual stationary or slowly varying propagation components. In contrast, downward-sloping traces, characterized by decreasing propagation delays, originate from vehicles approaching the measurement setup along the nearest traffic lane. The weaker upward-sloping traces are associated with vehicles traveling in the opposite direction along the more distant lane. Approximately 16 approaching and 10 receding vehicles can be distinguished during the analyzed recording. Some thicker or overlapping traces correspond to several vehicles passing in close succession. The slopes of these traces are related to the radial velocities of the moving vehicles, while their relative intensities depend on the corresponding scattering characteristics and propagation conditions.

Fig. 4 shows three example PDPs over the positive-delay range of 0–4 μs. Each profile is normalized to its maximum within the 0–1 μs interval used for RMS delay spread estimation. The estimated mean noise level in the late-delay region ($3.5 < \tau \leq 4$ μs) is −46.6 dB for the three illustrated profiles and −43.2 dB across all valid PDPs. These estimates provide a noise reference for the threshold sensitivity analysis. The estimated background levels indicate that the –40 dB cutoff remains above the mean background power for most PDPs, whereas the –50 dB cutoff may include noise-dominated samples.

To investigate the influence of PDP thresholding on the estimated temporal dispersion, three relative cutoff thresholds, $Th(\text{dB}) = \{-20, -30, -40\}$, were initially considered. For each PDP, the threshold was defined relative to its maximum power, and all components below the selected level were excluded from the delay spread calculation. The RMS delay spread, $\sigma_\tau$, was then evaluated using (1) for each consecutive PDP and threshold setting.

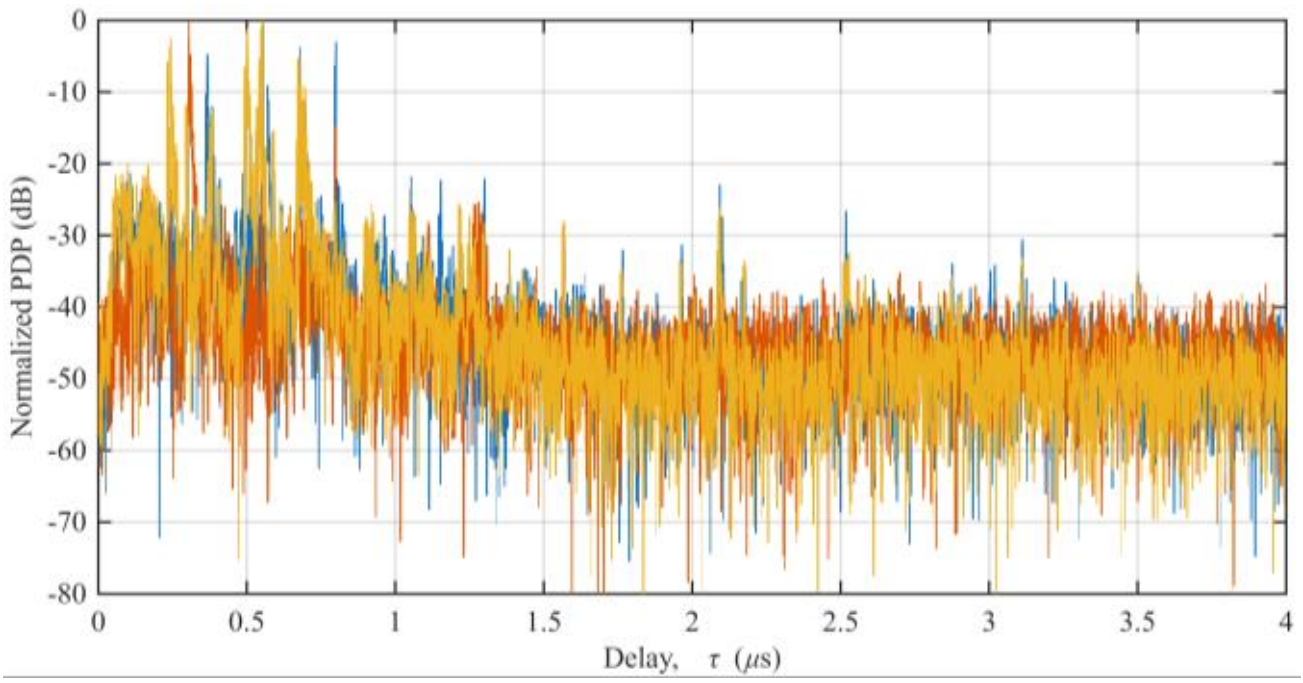

Fig. 4. Example normalized PDPs for $t$ = {0.225, 13.750, and 19.275} s.

Fig. 5 presents the resulting time variations of $\sigma_\tau$ over the 30 s observation interval. The yellow, red, and blue curves correspond to thresholds of –20, –30, and –40 dB, respectively. Although the three curves exhibit similar temporal trends, lower cutoff thresholds generally yield higher RMS delay spread values. This behavior results from the inclusion of weaker multipath components, which may contribute significantly to the second-order moment of the PDP, particularly when their delays differ substantially from the mean delay.

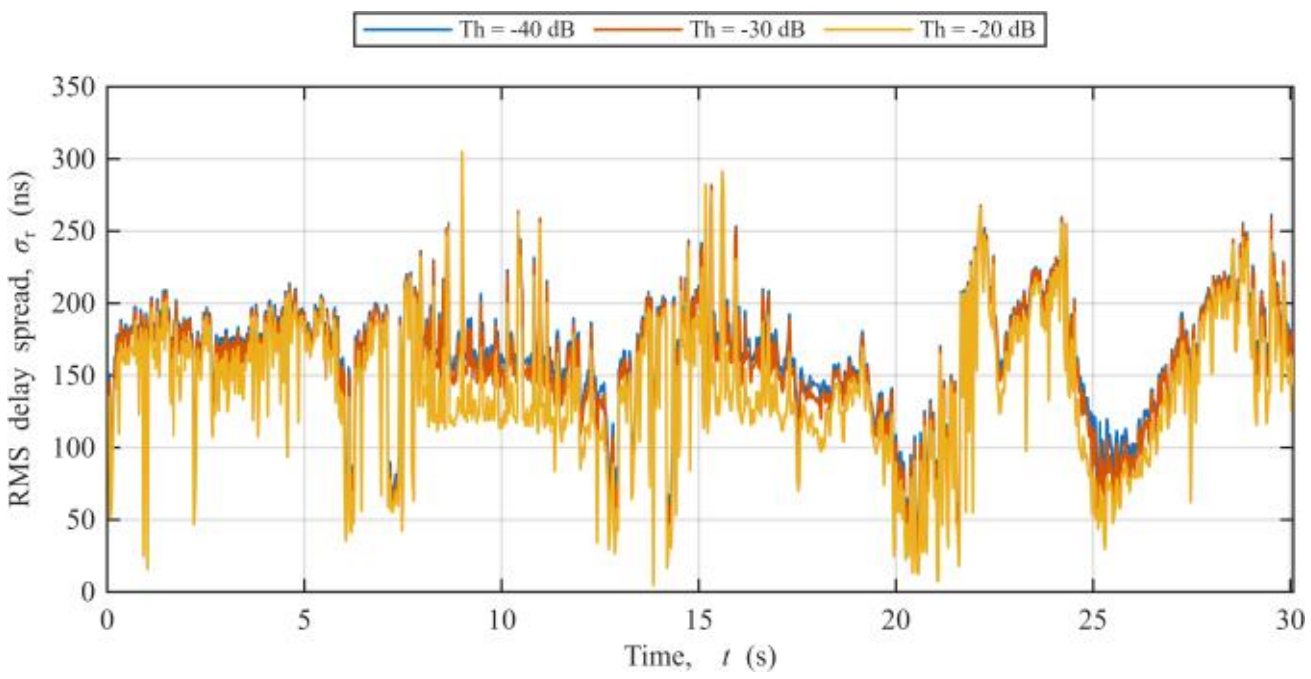

Fig. 5. Time-varying RMS delay spread at 33 GHz for three PDP cutoff thresholds.

The effect of threshold selection is more clearly illustrated by the empirical cumulative distribution functions (CDFs) of $\sigma_\tau$ presented in Fig. 6. The CDF obtained for $Th = -20$ dB is noticeably shifted toward lower delay spread values compared with those corresponding to –30 and –40 dB. The differences between the latter two distributions are considerably smaller, although still observable. These results indicate that the choice of cutoff threshold affects not only individual RMS delay spread estimates but also their statistical distribution.

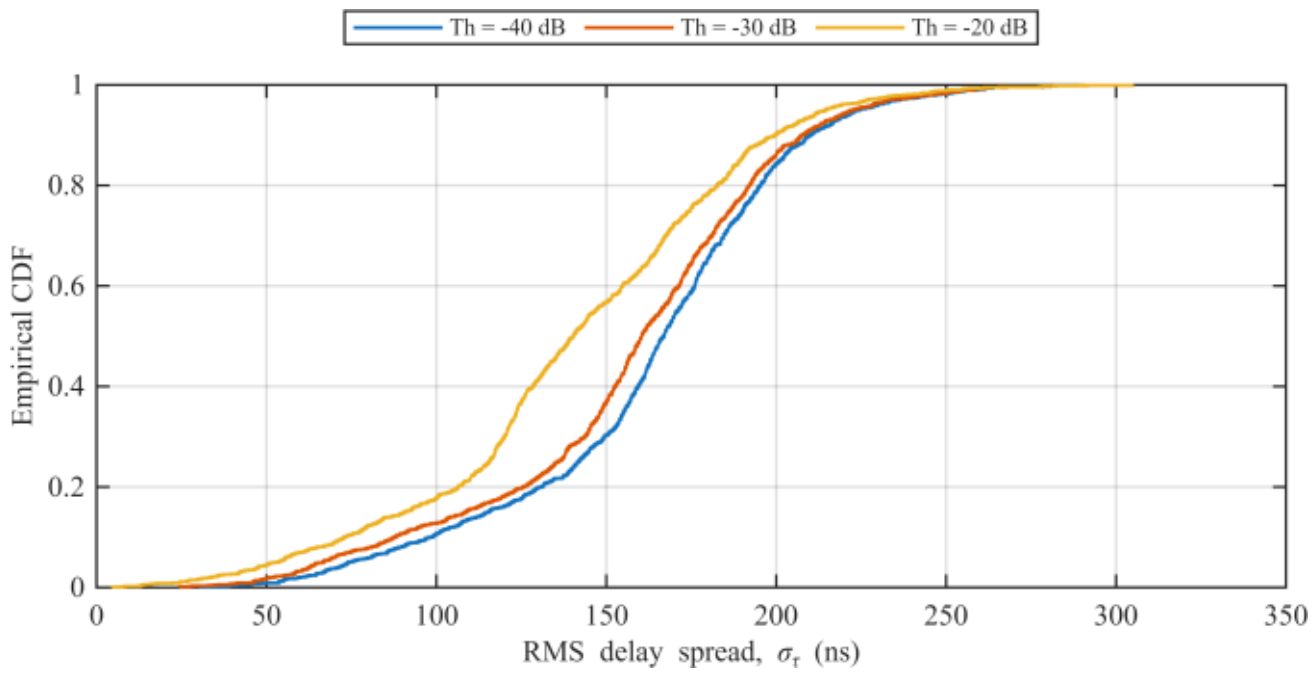

Fig. 6. Empirical CDFs of RMS delay spread for different PDP cutoff thresholds.

To further quantify this dependence, the median and P90 of the empirical RMS delay spread distributions were evaluated for cutoff thresholds ranging from –20 to –50 dB, with a step of –3 dB. The results are summarized in Fig. 7. As the threshold decreases from –20 to –50 dB, the median RMS delay spread increases from approximately 140 to 168 ns, whereas P90 increases from approximately 200 to 210 ns. The median approaches its maximum value for thresholds below approximately –41 dB, while P90 stabilizes at thresholds below approximately –35 dB.

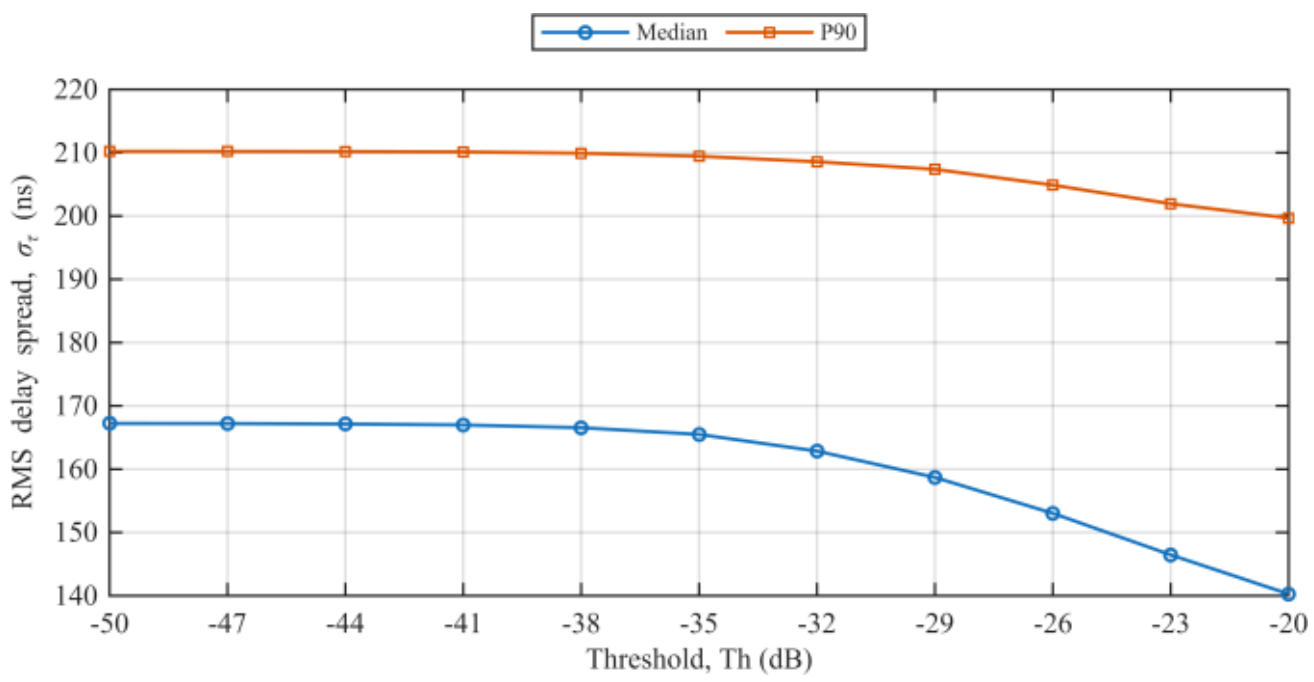

Fig. 7. Median and P90 of RMS delay spread versus PDP cutoff threshold.

The obtained results demonstrate that the selection of the relative PDP power threshold substantially affects RMS delay spread estimation in the investigated time-varying mmWave channel. The most pronounced differences are observed for thresholds between −20 and −30 dB, whereas lowering the threshold from −30 to −40 dB produces comparatively minor changes. For the considered measurement data, a threshold of approximately −40 dB can therefore be regarded as a practical convergence level for both statistics, as further threshold reduction produces negligible changes in the median and P90 of RMS delay spread.

The observed threshold dependence and convergence behavior are consistent with the findings reported by Pomianek *et al.* [6], who demonstrated the sensitivity of RMS delay spread to the considered PDP dynamic range. Moreover, the results support the methodological concerns raised by Schulpen *et al.* [4] regarding ambiguities in RMS delay spread estimation and the need for consistent thresholding criteria. Unlike these earlier investigations, the present study focuses on the temporal evolution and statistical convergence of RMS delay spread in a strongly dispersive, vehicle-induced time-varying channel at 33 GHz. Nevertheless, the identified convergence level should not be regarded as a universal threshold, since its applicability depends on the measurement dynamic range and noise floor characteristics, as also emphasized in [4], [5]. Consequently, RMS delay spread statistics obtained from different measurement campaigns should be compared only when the adopted PDP thresholding criteria are explicitly considered.

## IV. Conclusion

This paper investigated the impact of PDP thresholding on RMS delay spread estimation in a time-varying mmWave channel at 33 GHz. The analysis was based on empirical wideband measurements conducted in a dynamic road environment, where moving vehicles introduced significant variations in multipath propagation conditions.

The results demonstrated that the selection of the relative PDP power threshold substantially affects the estimated RMS delay spread and its statistical distribution. Decreasing the threshold from −20 to −50 dB increased the median RMS delay spread from approximately 140 to 168 ns, while the P90 increased from 200 to 210 ns. The most pronounced differences were observed between −20 and −30 dB, whereas further threshold reduction resulted in progressively smaller changes. The median and P90 approached stable values below approximately −41 and −35 dB, respectively, indicating that a threshold of −40 dB provides a practical convergence level for the investigated measurement data.

These findings confirm the importance of consistent PDP thresholding in empirical channel characterization, particularly when comparing RMS delay spread statistics obtained from different measurement campaigns. Although the identified convergence threshold is specific to the considered propagation scenario and measurement conditions, the presented approach provides a useful basis for assessing threshold sensitivity in other time-varying mmWave channels.

## Acknowledgment

This work was co-funded by the National Science Centre, Poland, under the OPUS call in the Weave programme under research project no. 2021/43/I/ST7/03294 acronym 'MubaMilWave'; by the Czech Science Foundation under project no. 23-04304L; by the Military University of Technology under project no. UGB 531-000128-W300-22/2026 "Analysis of the possibility of using modern technologies in communication systems and electronic warfare"; and by the MeitY, Government of India, through the Chips-to-Startup (C2S) program under project no. EE-9/2/2021-R&D-E.